\documentclass[aps,pre,twocolumn,showpacs,superscriptaddress]{revtex4}

\usepackage{graphicx}
\usepackage[tbtags]{amsmath}

\def\({\left(}
\def\){\right)}
\def\[{\left[}
\def\]{\right]}

\begin{document}
\title{Market behavior and performance of different strategy evaluation schemes}

\author{Yongjoo Baek}
\email{yjbaek@kaist.ac.kr} \affiliation {Department of Physics,
Korea Advanced Institute of Science
and Technology, Daejeon 305-701, Korea}

\author{Sang Hoon Lee}
\email[Present address: Department of Physics, Ume{\aa} University,
901 87 Ume{\aa}, Sweden]{}
\affiliation{Department of Physics, Korea Advanced Institute
of Science and Technology, Daejeon 305-701, Korea}

\author{Hawoong Jeong}
\email[Corresponding author: ]{hjeong@kaist.edu} \affiliation {Department of Physics,
Korea Advanced Institute of Science
and Technology, Daejeon 305-701, Korea}
\affiliation {Institute for the BioCentury, Korea Advanced Institute of Science
and Technology, Daejeon 305-701, Korea}
\date{\today}

\begin{abstract}
Strategy evaluation schemes are a crucial factor in any agent-based market model, as they determine the agents' strategy preferences and consequently their behavioral pattern. This study investigates how the strategy evaluation schemes adopted by agents affect their performance in conjunction with the market circumstances. We observe the performance of three strategy evaluation schemes, the history-dependent wealth game, the trend-opposing minority game, and the trend-following majority game, in a stock market where the price is exogenously determined. The price is either directly adopted from the real stock market indices or generated with a Markov chain of order $\le 2$. Each scheme's success is quantified by average wealth accumulated by the traders equipped with the scheme. The wealth game, as it learns from the history, shows relatively good performance unless the market is highly unpredictable. The majority game is successful in a trendy market dominated by long periods of sustained price increase or decrease. On the other hand, the minority game is suitable for a market with persistent zig-zag price patterns. We also discuss the consequence of implementing finite memory in the scoring processes of strategies. Our findings suggest under which market circumstances each evaluation scheme is appropriate for modeling the behavior of real market traders.
\end{abstract}


\maketitle

\section{Introduction} \label{sec:intro}
{\em Bounded rationality}~\cite{Simon1955} is now widely accepted as a fundamental aspect of human decision-making process. Insufficient information and cognitive limitations force people to rely on lessons of experience rather than reason out the optimal solution. Hence, especially in the fields of behavioral economics and econophysics, traditional assumptions of {\em perfectly rational} agents making {\em a priori} optimal decisions have been replaced by agents of limited cognitive capacity who follow rules of thumb empirically validated. An archetypical implementation of bounded rationality is found in the famous {\em El Farol Bar} problem~\cite{Arthur1994} and its simplified variant, the {\em minority game}~\cite{Challet1997,ChalletBook,CoolenBook,Savit1999,Johnson1999,Cavagna1999,Challet2000b,Marsili2001}. In the minority game, played by multiple agents for multiple turns, each player opts for one of two choices at each turn and wins the turn if the player is on the minority side. Since it is impossible to predict the choice of the other players, an optimal choice simply does not exist. Instead, each player forms a set of {\em strategies} which predict the behavior of the other players and advice the player which side to choose given the latest {\em choice pattern} of the other players. Using a {\em strategy evaluation scheme}, an agent compares the predictions of its strategies with the actual outcome of the game and evaluates the credibility of each strategy. At each turn, the advice of the most credible strategy is followed. In the standard minority game, strategies are evaluated in terms of their minority game score, {\em i.e.}, how often they correctly land on the minority side. Different strategy evaluation schemes may be used in other models of bounded rationality, but the structure of those models are essentially similar to the one outlined here.

The minority game has been readily applied in modeling financial markets~\cite{ChalletBook,Savit1999,Cavagna1999}. Its boundedly rational agents choosing from strategies bear some resemblance to real market participants. The fluctuating choice pattern shows features reminiscent of {\em stylized facts} of financial markets~\cite{Farmer1999}. Phase transition between symmetric and asymmetric regimes gives clues as to how markets self-organize themselves to be {\em marginally efficient}~\cite{YCZhang1999}. Yet whether the minority game faithfully captures the behavior of financial market speculators has been put under question~\cite{ChalletBook,Andersen2002,Challet2008}. One can observe that excess demand (supply) works to the advantage of sellers (buyers), thus favoring the minority, but this is merely a superficial similarity. Whether a strategy is successful is fully revealed only when an agent buys (short-sells) an item and sells (buys) it back {\em later} at a higher (lower) price. While the outcome of a single event determines the payoff in the minority game, in real markets we need at least two events separated in time to determine the payoff.

A variety of alternative models have been proposed to address this problem. One class of them assume that an agent buys (sells) an item and then immediately sells (buys) it back at the next time step. Marsili~\cite{Marsili2001} showed that if agents {\em expect} price changes in the adjacent time steps to be negatively correlated, the minority game payoff is justified. But if agents expect the price changes to be positively correlated, they should use the {\em majority game} payoff to evaluate their strategies. In that case, the strategies opting for the majority decision obtain higher scores. Since both expectations about price behavior are equally justifiable, we are led to an alternative market model where agents of both minority and majority expectations coexist. Meanwhile, starting from the same buy-today-and-sell-tomorrow assumption, other studies~\cite{Giardina2002,Andersen2002} derived the \$-game payoff which can be roughly considered a {\em time-delayed} version of the majority game payoff~\cite{payoff_comment}.

Another class of alternatives involves wealth, the total value of cash and assets in an agent's possession. Besides being a natural measure of success, wealth can be updated at each time step without considering two temporally separated events. Furthermore, past buy or sell decisions continue to affect the way score changes, since wealth fluctuates according to the value of financial items accumulated through time. This is not the case for the minority game or the majority game, since the score change is dependent only on the decision made in the previous turn. Hence agents evaluating their strategies using wealth-based payoff are far more history-dependent, and their collective behavior often leads to quite regular price behavior which may explain how different price trends are formed. Yeung {\em et al.}~\cite{CHYeung2008} performed a comprehensive study on a market model incorporating wealth as the measure of both an agent's success and a strategy's credibility, which they termed the {\em wealth game}.

All these different payoff schemes grew out of efforts to capture the characteristics of real financial markets more closely. Now it is natural to ask which ones are more relevant. Taking a Darwinian perspective, market participants are likely to use strategy evaluation schemes that are most beneficial for them, {\em i.e.}, best at accumulating wealth~\cite{Wiesinger2010}. Thus we may let agents with different strategy evaluation schemes participate in a market, and compare their performance on the basis of wealth. For example, Andersen {\em et al.}~\cite{Andersen2002} compared the wealth of the best minority game player with that of the worst in a market exclusively composed of minority game players, where the price is completely determined by the collective behavior of the minority game players. They found that the best player was actually poorer than the worst player in terms of wealth. This inconsistency implies that minority mechanism cannot dominate the {\em entire} market for long.

However, if minority game players account for only a small part of the market and are effectively decoupled from the price dynamics, there can be special situations when their strategy evaluation scheme proves profitable. In such cases, it is more suitable to follow the methodology set out by Yeung {\em et al.}~\cite{CHYeung2008}, which compares the average wealth achieved by different strategy evaluation schemes when the price data are {\em exogenously} given. Yeung {\em et al.} used price data taken from real financial markets, such as the Hang Seng Index (HSI). Since real market data cannot be directly controlled and their complexity evades any simple quantitative description, if we experiment with such data, it is hard to draw any general conclusions about the relation between the market trend and the corresponding suitable behavior patterns of agents. In order to clarify the relation {\em systematically}, we use artificial price data generated with a Markov process characterized by at most two parameters. Since we can now try various kinds of price behavior controlled by as few parameters as possible, we hope our results are better established and have more general implications.

This paper is organized as follows. First, we describe our model in Sec.~\ref{sec:model}. To get some intuition about the problem, we compare the performance of different strategy evaluation schemes in real markets in Sec.~\ref{sec:real}, using the Korea Composite Stock Price Index (KOSPI) and the HSI as price data. Conclusions drawn from this section are systematically verified in Sec.~\ref{sec:art}, using artificial price data generated by the Markov process. While our studies mainly concern strategy evaluation schemes with infinite score memory, whether introducing finite score memory changes the result is discussed in Sec.~\ref{sec:scorememory}. Finally, the summary of our results is presented in Sec.~\ref{sec:conclusions}.

\section{Model}
\label{sec:model}
We use a variant of the original wealth game model~\cite{CHYeung2008}, where the agents' buy or sell decisions have no influence on the price dynamics. This modification allows us to directly control the price dynamics, so that we can study the correlation between the price behavior and the profitability of each strategy evaluation scheme. Our ``exogenized'' model can be considered an approximation of the reality, if it concerns only a small fraction of the entire group of market participants. The situation is quite similar to the canonical ensemble in statistical mechanics, where we consider the temperature of the system a variable directly controlled by the external heat bath whose heat capacity is much larger than that of the system.

Consider $N$ agents participating in a stock market. As previously pointed out, they account for only a small part of the market. At time step $t$, agent $i$ decides whether to buy or sell a unit of stock, or to abstain from trade. Agent $i$'s each possible action is represented by $a_i\(t\)=\pm 1,~0$, respectively. Agent $i$'s {\it position} is the accumulation of the agent's past actions, written by
	\begin{equation} \label{eq:position}
		k_i\(t\)=\sum^{t-1}_{t'=0}a_i\(t'\).
	\end{equation}
This indicates the amount of stock the agent owns (if positive) or owes (if negative). Agent $i$'s wealth is the sum of its cash $c_i\(t\)$ and stock,
	\begin{equation} \label{eq:wealth}
		w_i\(t\)=c_i\(t\)+k_i\(t\)P\(t\),
	\end{equation}
where $P\(t\)$ is the price of a unit of stock. At each time step, the agent's action changes its cash by
	\begin{equation} \label{eq:cash}
		c_i\(t+1\)=c_i\(t\)-a_i\(t\)P\(t+1\).
	\end{equation}
Therefore the agent's wealth is updated by the rule
	\begin{equation} \label{eq:wealthupdate}
		w_i\(t+1\)=w_i\(t\)+k_i\(t\)\[P\(t+1\)-P\(t\)\].
	\end{equation}

The agents share an $m$-bit {\it market history}, which records the price increase (denoted by 1) and decrease (denoted by 0) for the latest $m$ time steps. Also each agent is provided with $s$ randomly drawn {\it strategies}. A strategy is a mapping from the set of $2^m$ possible market histories to the set of the agent's three possible actions (buy, sell, or abstain), the total number of possible strategies being $3^{2^m}$.

At every time step the agent updates the scores of its strategies according to a certain strategy evaluation scheme. The agent follows the suggestion of the highest-scoring strategy, although there is one exception to this rule. A suggestion that makes the agent's cash change sign from positive to negative, or decreases the cash that is already negative if followed is ignored and replaced with an abstention. This constraint, which we shall call the {\em non-negative cash constraint}, is implemented by introducing the position limitation
\begin{equation} \label{eq:maxpos}
	K_i\(t\) = \max \[\frac{w_i\(t\)}{P\(t\)},0\]
\end{equation}
so that any action increasing $|k_i| - K_i$ when $k_i > K_i$ or $k_i < -K_i$ is forbidden.

We consider three strategy evaluation schemes as classified by Yeung {\em et al.}~\cite{CHYeung2008} The score of strategy $\sigma$ at time step $t$ is denoted by $u_{\sigma}\(t\)$.

\begin{enumerate}
\renewcommand{\labelenumi}{(\roman{enumi})}
\item Wealth game (WG):
the score of a strategy is updated in a manner similar to the way an agent's wealth is updated. That is, the score of strategy $\sigma$ is updated by
	\begin{equation}
	u_{\sigma}\(t+1\)=u_{\sigma}\(t\)+k_{\sigma}\(t\)\[P\(t+1\)-P\(t\)\]
	\label{eq:wg}
	\end{equation}
where $k_\sigma\(t\)=\sum^{t-1}_{t'=0}a_\sigma\(t'\)$ is the {\em virtual position} of strategy $\sigma$. Now strategies are also subject to the non-negative cash constraint: any action $a_\sigma\(t\)$ of the strategy that makes the cash part of the strategy's score change sign from positive to negative, or decreases the cash part that is already negative, is replaced with an abstention. Note that while agent $i$'s wealth and its strategy $\sigma$'s score are updated in the same manner, they are {\em not} equal, since agent $i$'s action $a_i\(t\)$ and position $k_i\(t\)$ are different from strategy $\sigma$'s suggested action $a_\sigma\(t\)$ and virtual position $k_\sigma\(t\)$. We expect this position dependence to grant WG the strongest history dependence among the three schemes under consideration.

\item Minority game (MinG):
the score updating rule of MinG is given by
	\begin{equation}
	\label{eq:ming}
	u_{\sigma}\(t+1\)=u_{\sigma}\(t\)-a_{\sigma}\(t\)\[P\(t+1\)-P\(t\)\].
	\end{equation}
Trend-opposing strategies are favorably scored, as buying (selling) increases the score when the price falls (rises). The buying (selling) party can be roughly considered the minority in a bearish (bullish) market, so we can say MinG favors the minority. Hence the name of the scheme, despite the fact that the agents do not necessarily have to be in the minority in the exact sense of the term to gain profits. As pointed out by Marsili~\cite{Marsili2001}, users of this scheme believe that the immediate price trend will be reversed soon. Also they are short-sighted (or eager to forget about the past) in the sense that only the action of the previous turn, rather than the position, affects the change of score.

\item Majority game (MajG):
the score updating rule of MajG is given by
	\begin{equation}
	\label{eq:majg}
	u_{\sigma}\(t+1\)=u_{\sigma}\(t\)+a_{\sigma}\(t\)\[P\(t+1\)-P\(t\)\].
	\end{equation}
Trend-following strategies are favored, as buying (selling) increases the score when the price rises (falls). Just in the sense that MinG favors the minority, MajG favors the majority. MajG users expect that the current market trend would be sustained~\cite{Marsili2001}.
\end{enumerate}

\section{Performance in real markets} \label{sec:real}
We start by comparing the performance of the three strategy evaluation schemes in real financial markets. We use the closing price data of the HSI from December 31, 1986 to June 10, 2009 and the KOSPI from July 1, 1997 to June 10, 2009~\cite{Yahoo_finance}.

\subsection{Adaptation to market trends}
\begin{figure*}
		\centering
		\includegraphics[width=0.3\textwidth]{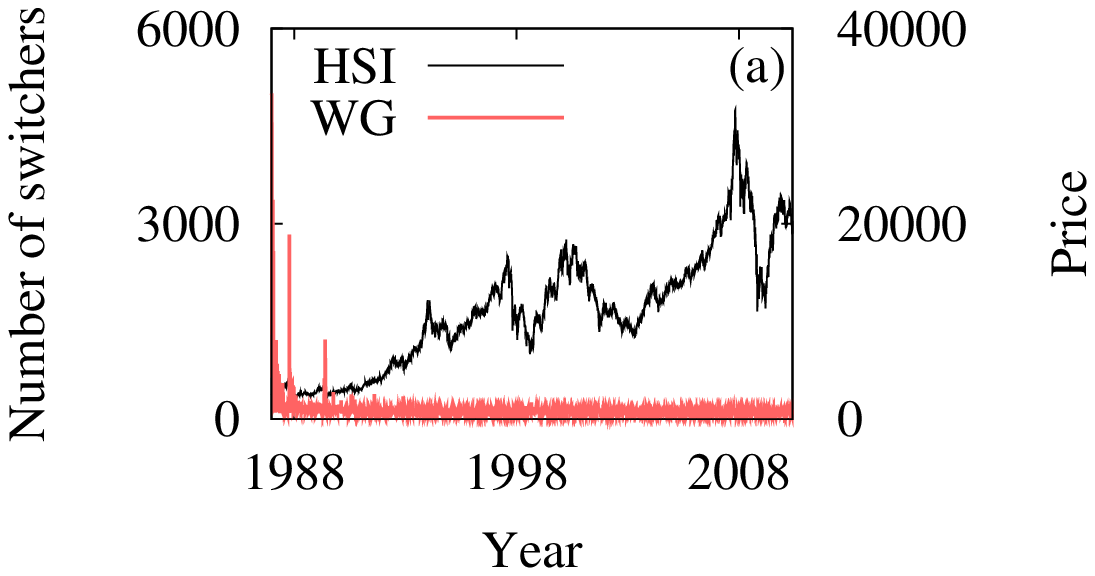}
		\includegraphics[width=0.3\textwidth]{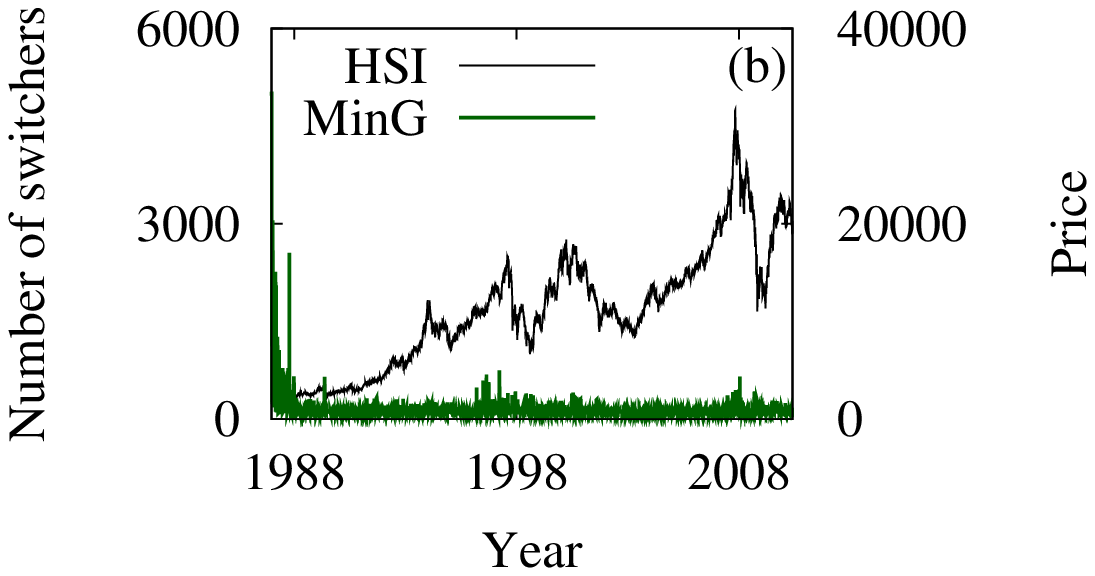}
		\includegraphics[width=0.3\textwidth]{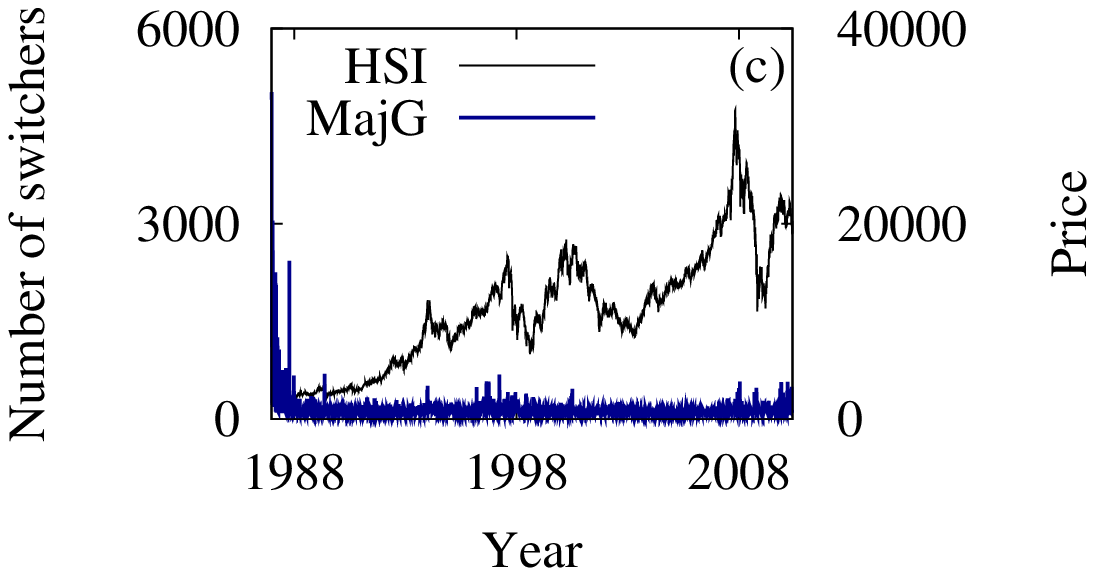} \\
		\includegraphics[width=0.3\textwidth]{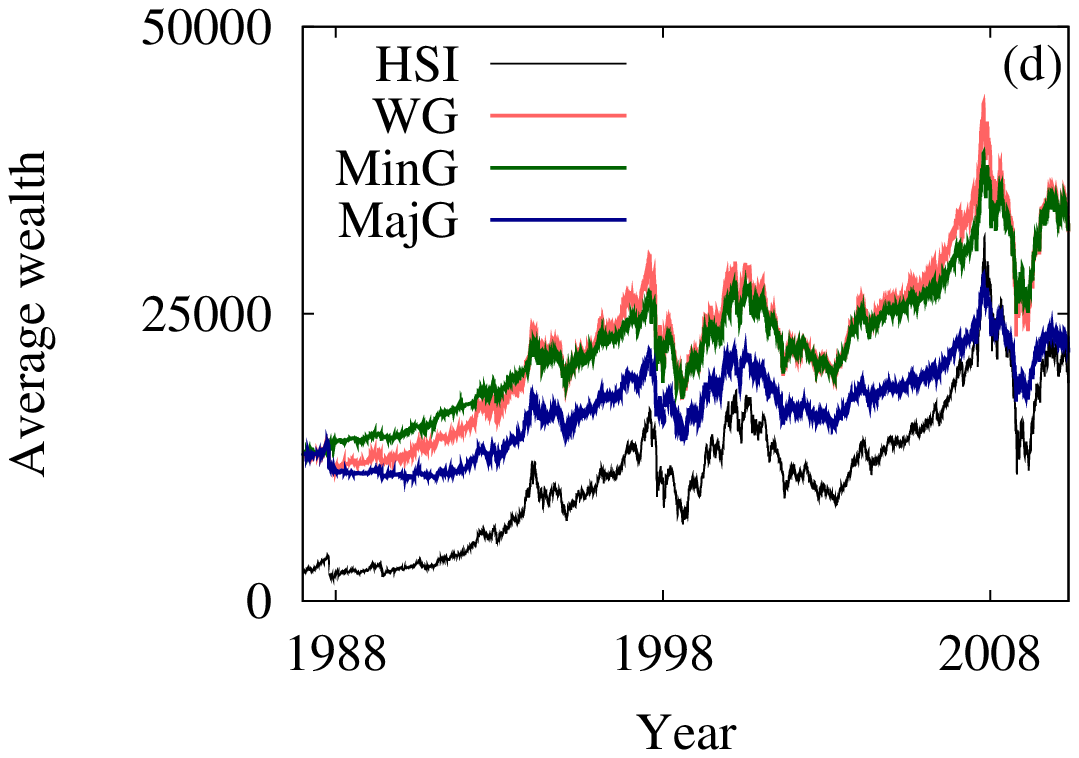}
		\includegraphics[width=0.3\textwidth]{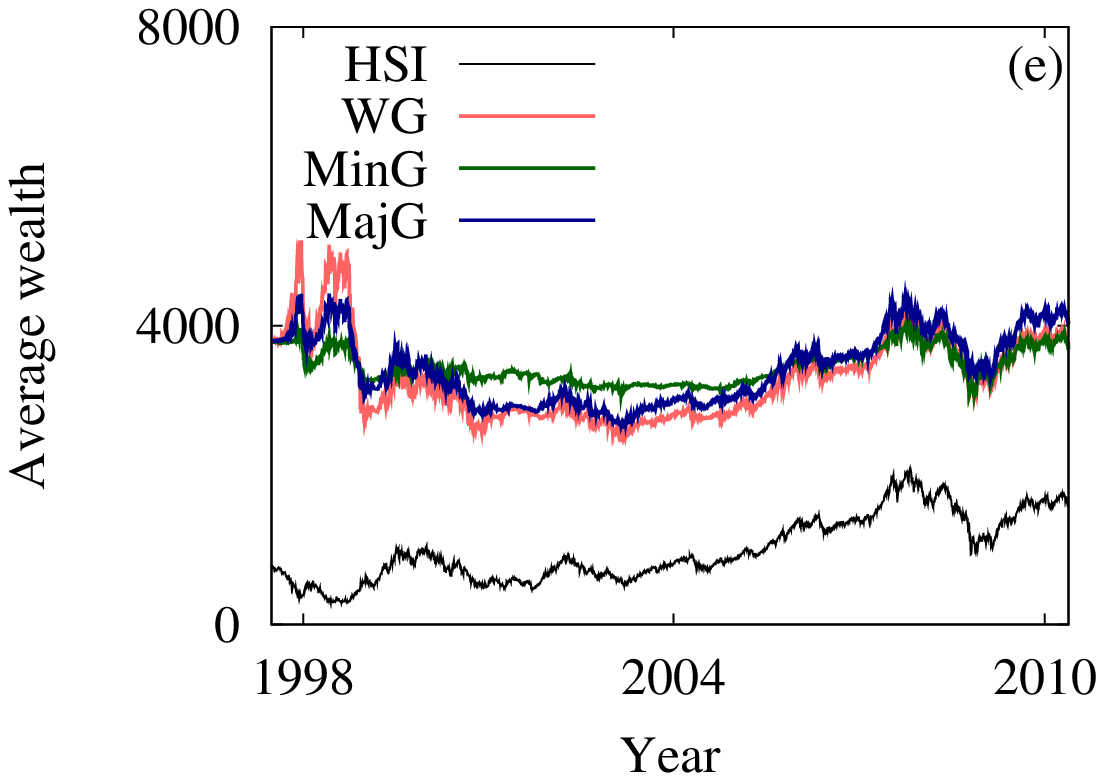}
		\caption{(Color online) The performance of the three strategy evaluation schemes are observed when the price is exogenously given by the real stock market indices: the HSI closing prices from December 31, 1986 to May 25, 2010 and the KOSPI closing prices from July 1, 1997 to May 25, 2010. Other market parameters are given by $N=10000$, $m=2$, $s=2$, and the initial wealth of each agent is set equal to five times the initial stock price to encourage activity. The number of strategy-switching agents is shown in the cases of (a) WG, (b) MinG, and (c) MajG. The average wealth gained by each strategy evaluation scheme is shown for both (d) HSI and (e) KOSPI. For reference, the stock market indices are shown along in black.}
		\label{fig:real}
	\end{figure*}

One measure of the agents' overall adaptation to the market trends is the number of strategy-switching agents. A large number of strategy-switching agents indicate that the agents are actively responding to changes in the market trend. As Figs.~\ref{fig:real}(a)--\ref{fig:real}(c) show, the peaks of the number of strategy switchers decrease in size, {\em i.e.}, later trend changes do not induce as much response from the agents as early trend changes do. Cumulative scoring of each strategy means that the score gap between strategies will broaden over time, which makes agents more reluctant to abandon previously successful strategies. Thus, the agents increasingly settle on one strategy.

Decay of peaks is the most rapid for WG. Position dependence of the WG wealth update rule quickly broadens the score gap between successful and unprofitable strategies, and this gap is not easily reduced unless a new market trend persists for a sufficiently long period. Thus WG agents tend to settle their preferred strategies early on and stick to it as long as possible. MinG and MajG agents show similar behavior, but their strategy preferences are not as clear as those of WG agents. Hence, MinG and MajG agents are more sensitive to trend changes.

\subsection{Effect of trend changes on agents' wealth}
In the previous subsection, we observed that WG agents quickly adapt to the initial market trend. They can optimize their positions for the initial trend and efficiently accumulate wealth. But since positions can be changed only by one at a time, they are very vulnerable to sudden trend changes. This is confirmed by the average wealth curves shown in Figs.~\ref{fig:real}(d) and \ref{fig:real}(e), especially the KOSPI during the 1997 Asian financial crisis. While WG agents quickly realized the benefits of short-selling and prospered during the initial market crash, their advantage turned into a trap when the market began to ``recover'' in 1998. Combination of large negative positions (built up by short-selling) and positive price changes meant WG agents were particularly hard hit by this sudden trend reversal. It took some time for WG agents to switch their strategies and reverse the sign of their positions, and for a while their average wealth was the lowest among the three schemes. This observation shows that while wealth is a convenient measure of success, it has its own shortcomings. An agent cannot liquidate its own assets all at once, so if a large portion of the agent's wealth comes from assets, the amount of wealth is largely at the mercy of price fluctuations.

MajG agents suffer similar difficulties in 1998, but their adaptation to the initial trend was not as complete as that of WG agents, {\em i.e.}, their positions were not sufficiently negative. Hence, their initial wealth gain on average was less than that of WG agents, but so was their loss due to the trend reversal.

On the other hand, MinG agents are always least affected by trend reversals. They always try to move against the market trend, and consequently their actions are severely restricted by the position constraint in Eq.~(\ref{eq:maxpos}), limiting their positions to near-zero region. Hence MinG agents may temporarily attain the highest average wealth after some trend changes. But they cannot keep the lead for long if the new trend turns out to be stable, as in the KOSPI where WG eventually catches up with MinG.

\subsection{Initial trend significance}
Agents are most flexible in the initial stage and grow increasingly conservative over time. As time passes, the gaps between the scores of strategies broaden and wealth gained or lost in the initial phase limits the freedom of the agents, both making it harder for agents to switch to a different strategy. Figure~\ref{fig:real}(d) shows that MinG agents are the second most successful in the HSI despite the general increase of price. This is because the initial trend, somewhat periodical increase and decrease of price, worked to MinG agents' advantage. This helped MinG agents attain good position and high average wealth early on, and MajG never overcame this initial disparity. A similar observation holds for the KOSPI, where WG was never able to make up for the advantage of MajG formed by the trend reversal in 1998.

This initial trend dependence may be seen as a weakness of the strategy evaluation schemes considered in this study. But while real markets show a complex mixture of {\em heterogeneous} trends, the strategy evaluation schemes are designed as a {\em simplified} model of the behavior of market participants. The extent to which those simplified agents can cope with ever-changing market trends is bound to be limited, and hence the lingering influence of the initial trend. Traders in real markets are certainly more adaptive than those model agents and initial trends would be less crucial for them. This is yet another reason why we experiment with artificially generated prices in the next section, so that the complexity of market trends is reduced and initial trend dependence becomes less significant.

In summary, WG agents quickly adapt to the initial market trend and draw the most profits from the trend. At the same time, however, WG agents suffer the most from sudden trend changes, while MG agents are least swayed by them. In addition, the initial trend is significant, since it affects the strategy preference and the freedom of choice for later periods.

\section{Performance in artificial markets} \label{sec:art}
\subsection{Simulation settings}
To clarify the relationship between the price pattern and the performance of strategy evaluation schemes remains, we extend our study to include artificially generated prices whose behavior can be described by a few parameters. Let $p_\uparrow (\mu)$ be the probability that the price will increase in the next time step given the latest $m$-bit market history $\mu$. At each time step the price can increase or decrease by 1. Starting from the random initial $\mu$ and the initial price $P(0) = 1000$, we generate all the subsequent price data.

\subsection{History-independent price behavior}
Consider the case when the price data are generated using only a single probability $p_\uparrow$, the history-independent probability of price increase. In other words, the price dynamics is a biased random walk. If $p_\uparrow$ is sufficiently larger or smaller than $1/2$, the price reliably increases or decreases from the beginning to the end. If $p_\uparrow \simeq 1/2$, the dynamics gets close to an unbiased random walk and the price behavior becomes unpredictable.

\begin{figure}
	\centering
	\includegraphics[width=0.9\columnwidth]{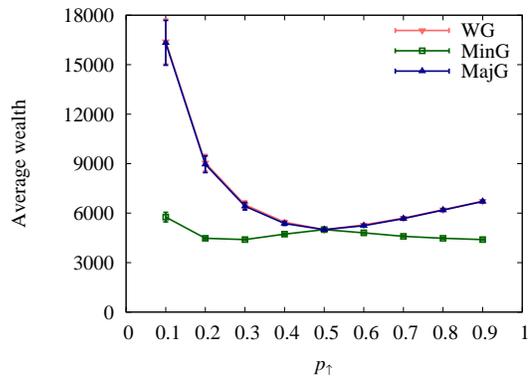}
	\caption{(Color online) The sample mean of the average wealth gained by each strategy evaluation scheme at the 1000th time step, measured for various values of the probability of price increase $p_\uparrow$ and averaged over 1000 samples. The other parameters are given by $N=10000$, $m=2$ and $s=2$. The error bars indicate standard deviation of the average wealth from the sample mean. Note that the lines are guide to the eyes.}
	\label{fig:mem0}
\end{figure}

Figure~\ref{fig:mem0} shows the average wealth of agents for each strategy evaluation scheme at the 1000th time step, for each value of $p_\uparrow$. As the market becomes more predictable ($p_\uparrow$ farther away from $1/2$), WG and MajG are similarly more successful than MinG. Higher predictability means that WG agents' lessons from history are more useful. Since there is only one probability involved, higher predictability is synonymous with stability of trends, a favorable condition for characteristic MajG agents. On the other hand, no strategy evaluation scheme is significantly better than the others when the market is unpredictable.

Note that the WG and MajG curves in Fig.~\ref{fig:mem0} are highly asymmetric. These curves indicate that agents are better off when $p_\uparrow$ is low rather than high, which is a natural consequence of our model. When $p_\uparrow$ is high, agents accumulate wealth by maintaining large positive positions. But as the agents run out of cash, the position constraint Eq.~(\ref{eq:maxpos}) prevents them from buying stock any further. Hence the average wealth increases only at a limited rate. When $p_\uparrow$ is low, successful agents quickly build up large negative positions. Such an agent's wealth has positive contribution from cash and negative contribution from the amount of sold stock. While the amount of cash keeps increasing, the fraction of stock in the agent's total wealth continues to decrease as the price falls.

Therefore, there is effectively no lower bound on the negative position, which leads to ever-accelerating wealth gain of agents. This result is against our intuition that bullish market is more profitable than bearish market, but it should be recalled that we are considering only a small fraction of the entire market participants with negligible influence on the price dynamics, who can freely buy or sell. Under realistic circumstances, agents cannot keep selling stock and build negative position, since they have trouble finding buyers.

Note that this asymmetry is also reflected in the real market simulation results shown in Fig.~\ref{fig:real}. Since the overall increment of the HSI is larger than that of the KOSPI during the observation period, the eventual ratio of the agents' average wealth to the stock index is much larger in case of the KOSPI than the HSI.

\subsection{Price generated by two-bit market history}
\begin{figure}
	\centering
 	\includegraphics[width=0.9\columnwidth]{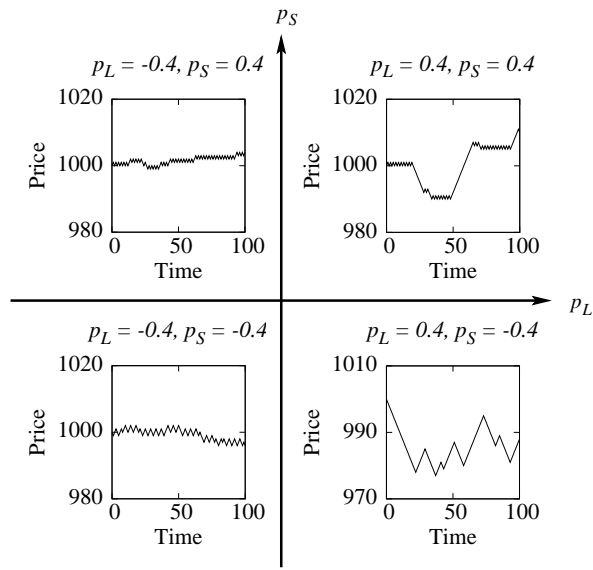}
 	\caption{(Color online) Insets show in purple examples of price data generated by different values of $p_L$ and $p_S$. Higher values of $p_L$ indicate greater likelihood for long-term trends, and higher values of $p_S$ indicate longer zig-zag oscillations.}
 	\label{fig:mem2_price}
\end{figure}

\begin{figure*}
 	\centering
 	\includegraphics[width=0.8\textwidth]{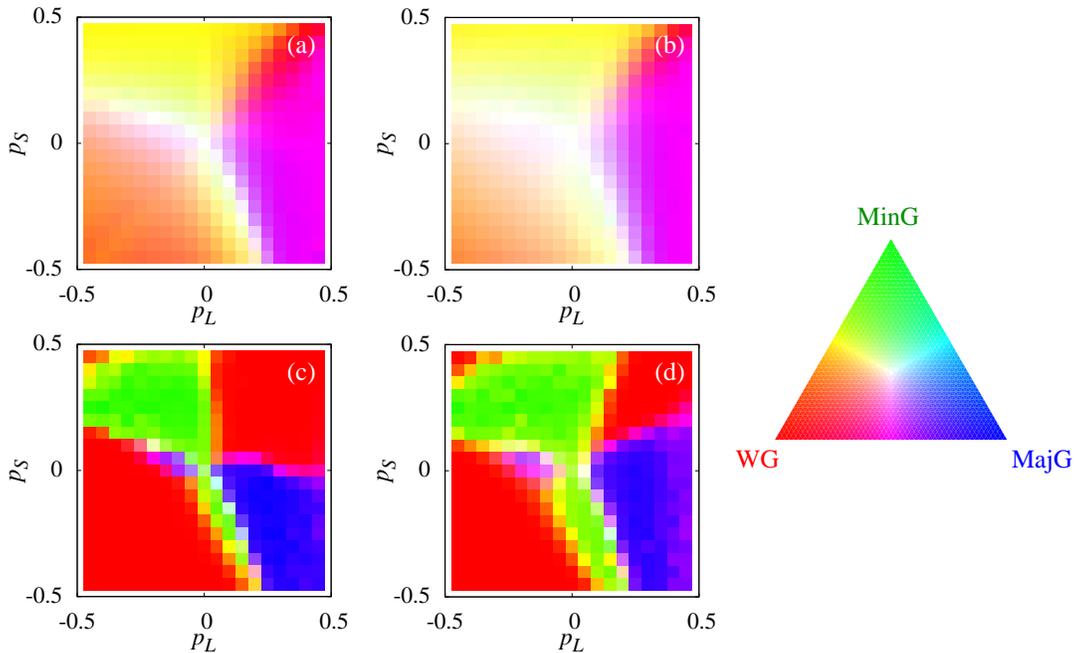}
 	\caption{(Color online) Diagrams showing performance of strategy evaluation schemes for different values of $p_L$ and $p_S$, with memory size $m$ varied. The upper two diagrams indicate the average wealth of agents using each strategy evaluation scheme, for (a) $m=2$ and (b) $m=3$. The lower two diagrams represent the chance of each evaluation scheme achieving the highest average wealth, for (c) $m=2$ and (d) $m=3$. Other parameters are given by $N=10000$ and $s=2$, with the average wealth measured at the 5000th time step. As the ternary plot on the right shows, superior performance is indicated by greater weight put on the scheme's representative color. We use the average wealth of agents and the chance of attaining the highest average wealth to visualize the relative performance of a strategy evaluation scheme. Note that the region where MinG or MajG is successful increases as $m$ is increased from $2$ to $4$. We observed the same trend when $m$ is further increased to $10$.}
 	\label{fig:mem2}
\end{figure*}

Now we generate the price data using four probability parameters, $p_\uparrow \left(\downarrow\downarrow\right)$, $p_\uparrow \left(\downarrow\uparrow\right)$, $p_\uparrow \left(\uparrow\downarrow\right)$, and $p_\uparrow \left(\uparrow\uparrow\right)$. Each parameter corresponds to the probability of price increase given the latest two-bit market history, where the direction of each arrow represents the direction of price change. In this case, we can say the price is generated by a Markov chain of order two.

Previous simulations show that long-term price increase or decrease favors WG and MajG over MinG. Now we shall only consider the cases when general price increase or decrease from the beginning to the end is suppressed. For this we introduce two new constraints,

\begin{equation}
	p_\uparrow \(\downarrow\downarrow\) + p_\uparrow \(\uparrow\uparrow\) = p_\uparrow \(\downarrow\uparrow\) + p_\uparrow \(\uparrow\downarrow\) = 1.
\label{prob_constraint}
\end{equation}

This reduces the number of parameters to two. We define the long-term parameter $p_L$ and the short-term parameter $p_S$ by

\begin{equation}
\begin{split}
	p_L &\equiv p_\uparrow \(\uparrow\uparrow\) - 0.5 = 0.5 - p_\uparrow \(\downarrow\downarrow\), \\
	p_S &\equiv p_\uparrow \(\uparrow\downarrow\) - 0.5 = 0.5 - p_\uparrow \(\downarrow\uparrow\).
\end{split}
\end{equation}

If $p_L > 0$, price is likely to increase or decrease for three consecutive time steps or longer. If $p_L < 0$, price increase or decrease is not likely to continue for more than two time steps. Thus $p_L$ controls the likelihood of long-term price trends. Meanwhile, if $p_S > 0$, price is likely to show sustained zig-zag oscillations of period two. If $p_S < 0$, price increase or decrease is likely to continue for at least two time steps. Hence $p_S$ controls how rapid price oscillations are likely to be.

Examples of generated price data for different values of $p_L$ and $p_S$ are shown in Fig.~\ref{fig:mem2_price}. Figure~\ref{fig:mem2} visualizes relative performance of the three evaluation schemes, in terms of the average wealth and the chance of achieving the highest average wealth. We can simplify our observations in terms of four extreme cases.

\begin{enumerate}
\renewcommand{\labelenumi}{(\roman{enumi})}
	\item $p_L > 0$ and $p_S > 0$ \\
	The price behavior is dominated by long-term trends with intermittent zig-zag oscillations, as illustrated by the $p_L = 0.4$ and $p_S = 0.4$ case shown in Fig.~\ref{fig:mem2_price}. WG and MajG outperform MinG since trend-following strategies are more suitable for a trendy market. However, zig-zag oscillations are also likely to persist, which is disadvantageous for MajG. Hence, WG tends to be more successful than MajG.
	\item $p_L > 0$ and $p_S < 0$ \\
	The price behavior is completely dominated by long-term trends without zig-zag oscillations, as illustrated by the $p_L = 0.4$ and $p_S = -0.4$ case shown in Fig.~\ref{fig:mem2_price}. WG and MajG are more successful than MinG, with MajG closely in the lead. It is not obvious why MajG should be better than WG, but note that there are abrupt trend reversals in the price pattern. As pointed out in Sec.~\ref{sec:real}, WG is particularly prone to initial trend reversals. Thus WG agents start with a slight disadvantage, which they find difficult to make up for later even if they eventually adapt themselves to the market.
	\item $p_L < 0$ and $p_S > 0$ \\
Long-term trends are suppressed and zig-zag oscillations of period 2 ($\uparrow\downarrow\uparrow\downarrow\ldots$, for example) become dominant, as illustrated by the $p_L = -0.4$ and $p_S = 0.4$ case shown in Fig.~\ref{fig:mem2_price}. MinG is the most successful since its fundamentalistic expectations turn out to be correct. WG still adapts well to the price behavior and its average wealth is only slightly less than that of MinG. MajG gets the lowest average wealth due to its chartistic nature.
	\item $p_L < 0$ and $p_S < 0$ \\
Zig-zag oscillations of period 4 ($\uparrow\uparrow\downarrow\downarrow\ldots$, for example) become dominant, as illustrated by the $p_L = -0.4$ and $p_S = -0.4$ case shown in Fig.~\ref{fig:mem2_price}. This condition is still more favorable for MinG than for MajG, but the advantage of MinG is not as strong as in the case $p_S > 0$. Thus, WG outperforms MinG.
\end{enumerate}

It should be noted that although MajG tends to show the worst performance for $p_L < 0$, there is an exception when $p_S \simeq 0$. Whenever $p_S$ is far from zero while $p_L < 0$, the price dynamics is dominated by regular zig-zag price oscillations. But if $p_S \simeq 0$, regularity of the oscillations is broken, giving some chance for MajG while impairing the performance of MinG.

Even if we change the values of the number of agents $N$ and the length of market history $m$ to check the stability of our results, qualitatively the same results as explained above are observed. Interestingly, the area of the $p_L$ -- $p_S$ diagram in which MinG or MajG is the most successful increases as $m$ is increased, as shown in Fig.~\ref{fig:mem2}. This is not because greater memory $m$ enhances the performance of MajG and MinG. Rather, all three schemes tend to show worse performance for greater $m$, with the impairment being the severest for WG. We suspect that when $m > 2$, agents are trying hard to find some spurious causal relations between market history and price behavior, which can cause undesirable inefficiency.

The price dynamics is dependent only upon the latest two time steps, but agents are considering longer time spans to make their decisions. Thus they are likely to make false conclusions about the market trends, which deteriorates their average wealth. Also greater $m$ leads to rapidly growing number of possible strategies ($= 3^{2^m}$), but we have fixed the number of strategies available for each agent to two. This further hinders the agents from making correct decisions, since only a few agents would be given strategies suitable for the market trends. Since WG has the strongest dependence on history, they suffer the heaviest loss from these problems.

In most cases, WG manages to be the most successful scheme, otherwise closely follows the scheme of the highest average wealth. Although they may suffer temporarily from trend changes, eventually they learn to make up for their loss as long as the market shows sufficient predictability. Even if the market totally lacks predictability, all schemes show similar performance, so even in such cases we cannot say WG is worse than other schemes. We can say WG is the most ``versatile'' among the three schemes.

With this concept of Markov chain, we are able to {\em measure} the probabilities of each movement $p_\uparrow \(\downarrow\downarrow\)$, $p_\uparrow \(\downarrow\uparrow\)$, $p_\uparrow \(\uparrow\downarrow\)$, and $p_\uparrow \(\uparrow\uparrow\)$ for the real stock market data, in retrospect. If the probabilities are measured for each consecutive two-step and averaged for the entire observation period, it turns out that for the HSI, $p_\uparrow \(\downarrow\downarrow\)=0.52$, $p_\uparrow \(\downarrow\uparrow\)=0.53$, $p_\uparrow \(\uparrow\downarrow\)=0.50$, and $p_\uparrow \(\uparrow\uparrow\)=0.53$. Similar values are observed for the KOSPI with $p_\uparrow \(\downarrow\downarrow\)=0.51$, $p_\uparrow \(\downarrow\uparrow\)=0.54$, $p_\uparrow \(\uparrow\downarrow\)=0.53$, and $p_\uparrow \(\uparrow\uparrow\)=0.56$. Note that the previous constraints in Eq.~(\ref{prob_constraint}) do not exactly hold in real data~\cite{real_Markov}. However, all the probabilities are slightly above $0.5$, $p_\uparrow \(\uparrow\uparrow\) \gtrsim p_\uparrow \(\downarrow\downarrow\)$, and $p_\uparrow \(\downarrow\uparrow\) \gtrsim p_\uparrow \(\uparrow\downarrow\)$, which may give slight overall advantage to WG and MajG according to Fig.~\ref{fig:mem2}(a) ($m = 2$ case), reasonably consistent with our observation.

\section{Finite Score Memory} \label{sec:scorememory}
Thus far, we have assumed that the scores of the strategies have infinite memory. That is, a strategy's suggestions made in the past and at present equally contribute to its score. As time passes, the sheer size of the past experience overwhelms contributions from the recent experience. Thus agents grow increasingly unresponsive to trend changes, as the decrease of strategy switchers in Figs.~\ref{fig:real}(a)--\ref{fig:real}(c) clearly indicates. For the agents to stay highly adaptive to market changes, our model should implement finite score memory so that the emphasis is put upon the ``present state of affairs.''

Finite score memory can be implemented by evaluating the performance of a strategy in terms of the score accumulated over the latest $T$ time steps~\cite{Lamper2001, rate_const}. If $u_\sigma\(t\)$ denotes the infinite-memory score of a strategy as originally defined by one of the three evaluation schemes explained in Sec.~\ref{sec:model}, we define the corresponding finite-memory score $u_\sigma^T\(t\)$ by
\begin{equation}
u_\sigma^T \(t\) = u_\sigma \(t\) - u_\sigma \(t-T\).
\end{equation}
It is straightforward to apply the above formula to MinG, MajG, and WG. For convenience, let these finite-score-memory variants of the original three strategy evaluation schemes be called the delta minority game (DMinG), the delta majority game (DMajG), and the delta wealth game (DWG).

A note of caution is in order in the case of DWG. The virtual wealth $u_\sigma \(t\)$ should obey the non-negative cash constraint explained in Sec.~\ref{sec:model}, if $u_\sigma^T \(t\)$ is to be interpreted as the {\em change} of virtual wealth. Since the score $u_\sigma^T \(t\)$ is indirectly subject to the constraint through $u_\sigma \(t\)$, the DWG score is still under the influence of the entire history. In this sense, the score memory of DWG is not truly finite.

\subsection{Real markets}

\begin{figure*}
 	\centering
 	\includegraphics[width=0.45\textwidth]{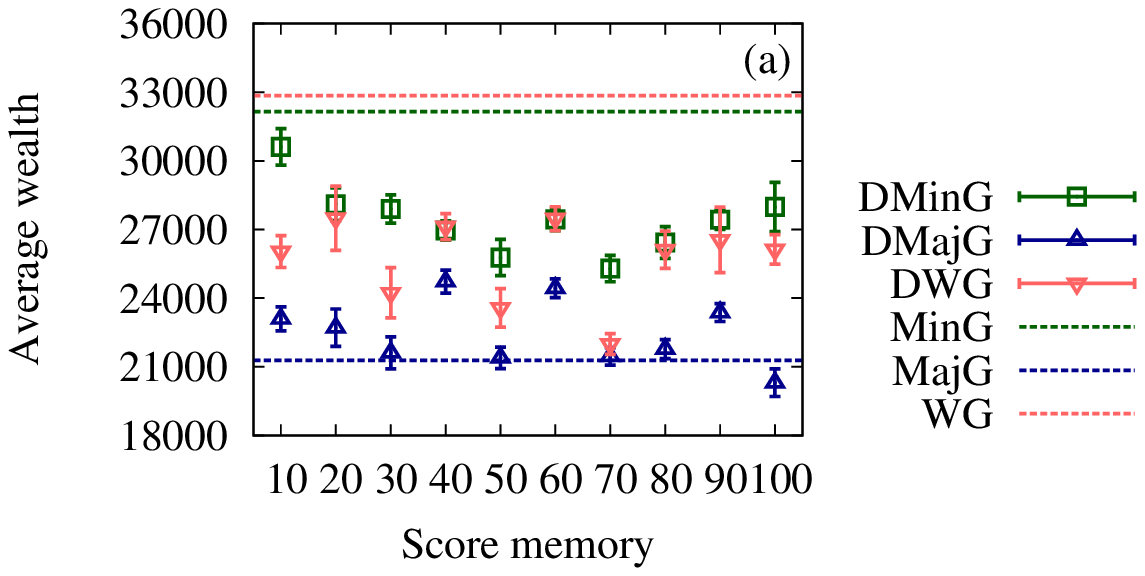}
	\includegraphics[width=0.45\textwidth]{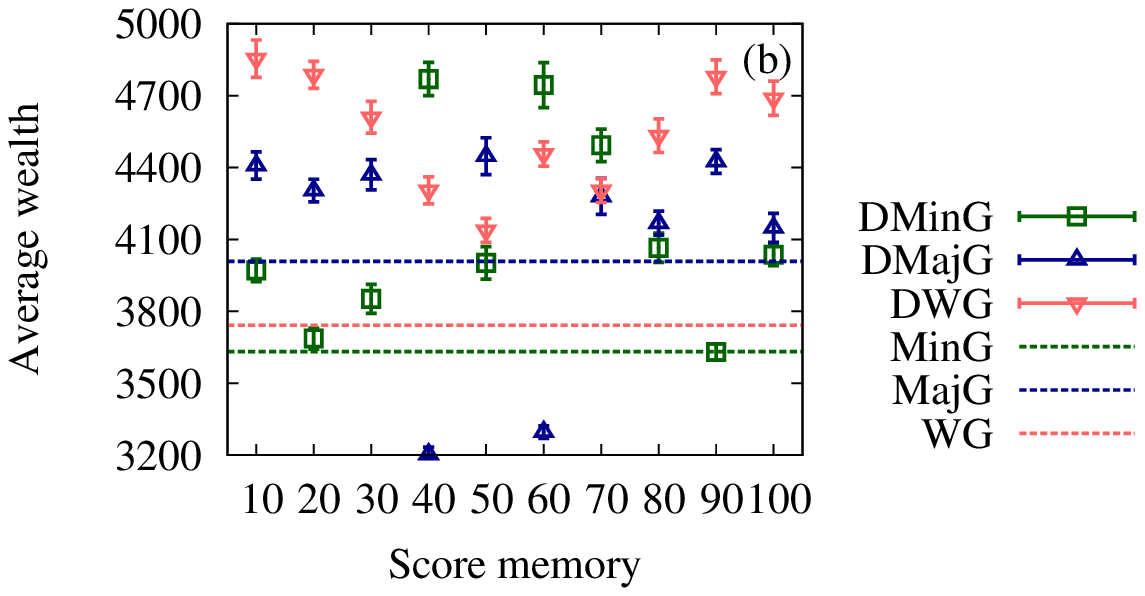}
 	\caption{(Color online) The data points represent the average wealth gained by each strategy evaluation scheme shown for different sizes of score memory $T$, while the dashed lines correspond to the average wealth obtained by the original infinite-score-memory evaluation schemes. The average wealth is measured at the end of the observation period, which is (a) from December 31, 1986 to May 25, 2010 for the HSI closing prices and (b) from July 1, 1997 to May 25, 2010 for the KOSPI closing prices. Parameters other than the score memory are given by $N=10000$, $m=2$, $s=2$, and an agent's initial wealth is set equal to five times the initial price.}
 	\label{fig:finscrmemreal}
\end{figure*}

Figure~\ref{fig:finscrmemreal} shows the average wealth achieved by DMinG, DMajG, and DWG in real markets (KOSPI and HSI) on the last day of the period of observation, for different score memory sizes $T$. The performance of each strategy evaluation scheme strongly depends on the score memory size, especially in the case of KOSPI where the overall increase of price is much less dominant. While it is hard to find any general tendencies in the fluctuating average wealth curves, they do suggest that the real market behavior is a complex mixture of market trends with different time scales.

\subsection{Artificial markets}

\begin{figure*}
 	\centering
 	\includegraphics[width=0.8\textwidth]{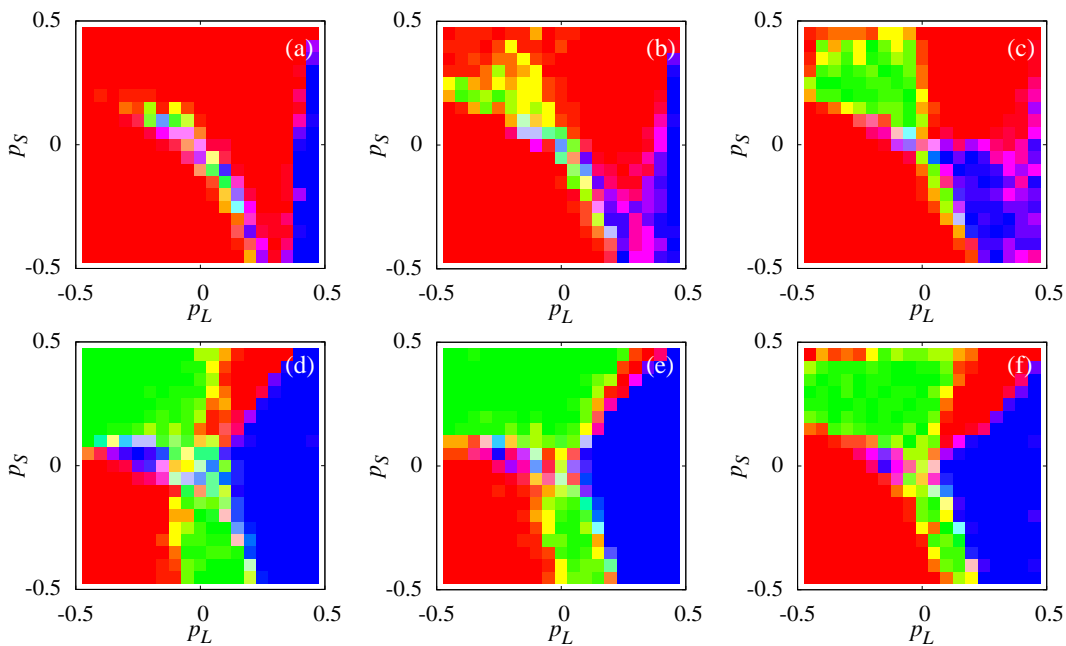}
 	\caption{(Color online) Each strategy evaluation scheme's chance of reaching the highest average wealth are shown for different values of the score memory size $T$, in an artificial market whose price is generated by a Markov chain of order 2. Refer to the ternary plot in Fig.~\ref{fig:mem2} for the meaning of each color. The upper three diagrams compare the performance of the original WG, DMinG, and DMajG for (a) $T=10$, (b) $T=100$, and (c) $T=1000$. The lower three diagrams compare the performance of DWG, DMinG, and DMajG for (d) $T=10$, (e) $T=100$, and (f) $T=1000$. Other parameters are given by $N=10000$, $s=2$, $m=2$ with the average wealth measured at the 5000th time step.}
 	\label{fig:finscrmemart}
\end{figure*}

Let us first consider the performance of WG, DMinG, and DMajG in an artificial market where the price is generated by biased random walk. We can observe that finite score memory does not make much difference in this case: DMajG exhibits the worst performance among the three except for the unpredictable case $p_{\uparrow} \sim 0.5$, while DWG and DMajG are almost equally successful, in a manner very similar to the result shown in Fig.~\ref{fig:mem0}. Finite score memory is intended to improve the performance of agents when the market trends change from time to time, but biased random walk generates very stable market trends. Thus score memory is not so relevant in this case.

If the price is generated by a Markov chain of order 2 subject to the constraint Eq.~(\ref{prob_constraint}), then finite score memory causes some visible changes, as shown in Fig.~\ref{fig:finscrmemart}. Figures~\ref{fig:finscrmemart}(a)--\ref{fig:finscrmemart}(c) show the relative performances of DMinG, DMajG, and the original WG. The original WG, whose score memory is fixed to be infinite, is unaffected by the score memory size $T$. Thus Figs.~\ref{fig:finscrmemart}(a)--\ref{fig:finscrmemart}(c) show how the {\em absolute} performances of DMinG and DMajG are affected by score memory. Note that the performances of DMinG and DMajG are worse than their infinite score memory counterparts, as indicaed by the greater area in the $p_L$ - $p_S$ diagram dominated by the original WG. As the score memory is increased, the shape of the $p_L$ - $p_S$ diagram approaches the infinite score memory result shown in Fig.~\ref{fig:mem2}. If we compare the performances of DMinG, DMajG, and DWG as shown in Figs.~\ref{fig:finscrmemart}(d)--\ref{fig:finscrmemart}(f), we observe that the performance of DWG is worse than that of the original WG, but again approaches the original infinite score memory result as $T$ becomes large.

Our observations indicate that finite score memory does not improve the agents' performance if the price is generated by a Markov process of order $\le 2$, while keeping the qualitative features of the observations made in the infinite score memory cases. The price series thus generated exhibit simple and steady market trends, while finite score memory is intended for more complex market trends. If the market trend is steady, agents would better recognize the trend if they observe it for a sufficiently long time rather than throw away their distant past experiences.

\section{Summary and conclusions} \label{sec:conclusions}
We have studied average performances of three strategy evaluation schemes, WG, MinG, and MajG, in a market whose price dynamics is exogenously determined by real market data or the Markov process. We believe that our simplified version of WG effectively captures the behavior of a small portion of traders in a large stock market, and helps determine which evaluation scheme is useful for individual traders given a particular market situation. We can expect that individuals whose influence on the market is negligible will use tests similar to the one presented in this paper, and choose the most successful scheme to evaluate their strategies. Thus we can tell which scheme most suitably describes the behavior of agents under different market circumstances. While incorporation of finite score memory does make a difference in real markets, the basic conclusions drawn from the artificial market observations are qualitatively still valid.

Note that MajG is most successful when the price pattern is completely dominated by long-term trends, while MinG is the best choice if the price tends to show rapid oscillations. In other words, these schemes prosper when their expectations are fulfilled. Combined with Marsili's observation~\cite{Marsili2001} that the price behavior follows the expectations of either MinG or MajG depending on which side is more dominant in the market, we get an idea of how certain price patterns maintain themselves through positive feedback. For instance, a bubble can maintain itself because long-term trends make MajG traders a dominant force in the market and MajG traders' expectations fulfill themselves. Whether we can model some {\em negative} counterpart of this feedback mechanism would be an interesting issue for further studies.

We used average wealth to assess the viability of each evaluation scheme, which at first glance may seem biased toward WG. But as our simulation results show, other schemes can be more successful than WG depending on the market behavior, even though wealth is the only measure of success in our model. Trend reversals and lingering influence of initial market trends are the main reasons why the performance of WG is impaired. WG agents are especially adept at building up positions optimal for the given price pattern, which means that a large portion of their wealth comes from their assets. Pattern changes can be too rapid for WG agents to follow up by moving to a new optimal position, and in such cases the agents are simply at the mercy of the market's whim.

Still, we cannot ignore the advantages of WG. Its main strength lies in its versatility, as shown by tests using generated price data. WG agents learns from the history, so they eventually adapt to any market trends and make up for the losses from initial ``mistakes.'' Thus, we can justify using WG to model the behavior of traders in financial markets, whenever there is some degree of predictability and persistence in price patterns.

\begin{acknowledgments}
This work was supported by NAP of Korea Research Council of Fundamental Science \& Technology (S.H.L.) and Basic Science Research Program through the NRF of Korea funded by the Ministry of Education, Science and Technology (Y.B. and H.J., 2009-0087691). Also Y.B. greatly thanks Professor K. Y. Michael Wong and his colleagues for their support and research guidance on the wealth game model and the real market data analysis,
during his visit to Hong Kong University of Science and Technology.
\end{acknowledgments}

\end{document}